\documentclass[12pt]{article}
\usepackage{a4wide}
\usepackage{amssymb,amsmath,amsthm,amscd}
\usepackage{graphicx}
\usepackage{enumerate}

\newcommand{\md}{{\mathrm d}}

\begin{document}


{\renewcommand{\thefootnote}{\fnsymbol{footnote}}
\begin{center}
{\LARGE  Propagators in Polymer Quantum Mechanics}\\

\vspace{1.5em}
Ernesto Flores-Gonz\'alez,\footnote{e-mail address: {\tt eflores@xanum.uam.mx}}
Hugo A. Morales-T\'ecotl,\footnote{e-mail address: {\tt hugo@xanum.uam.mx}}
and Juan D. Reyes\footnote{e-mail address: {\tt jdrp75@gmail.com }}
\\
\vspace{1em}
Departamento de F\'isica, Universidad Aut\'onoma Metropolitana-Iztapalapa, \\ 
San Rafael Atlixco 186, M\'exico D.F. 09340, M\'exico.
\end{center}
}
\setcounter{footnote}{0}


 


 

\begin{abstract}

Polymer Quantum Mechanics is based on some of the techniques used in the loop quantization of gravity that are adapted to describe systems possessing a finite number of degrees of freedom. It has been used in two ways: on one hand it has been used to represent some aspects of the loop quantization in a simpler context, and, on the other, it has been applied to each of the infinite  mechanical modes of other systems. Indeed, this polymer approach was recently implemented for the free scalar field propagator. In this work we compute the polymer propagators of the free particle and a particle in a box; amusingly, just as in the non polymeric case, the one of the particle in a box may be computed also from that of the free particle using the method of images. We verify the propagators hereby obtained satisfy standard properties such as: consistency with initial conditions, composition and Green's function character.  Furthermore they are also shown to reduce to the usual Schr\"odinger propagators in the limit of small parameter $\mu_0$, the length scale introduced in the polymer dynamics and which plays a role analog of that of Planck length in Quantum Gravity.

\end{abstract}






\section{Introduction}
Polymer quantum mechanics \cite{Ashtekar,Corichi1} is the theory obtained by quantizing a mechanical system adapting the techniques used in Loop Quantum Gravity (LQG). The latter is a background independent canonical quantization of General Relativity and a candidate theory of quantum gravity \cite{Thiemann,LQG}.


The Hilbert space and basic operators arrived at by applying the loop recipes in the case of finite degrees of freedom, constitutes a known example of  a so called singular representation of the canonical commutation relations of quantum mechanical systems \cite{Bratelli}.  As a singular representation, the polymer description is unitarily inequivalent to the Schr\"odinger representation \cite{KadisonRingrose1997}. This is to say that not only do the basic mathematical objects of the theory (state vectors, inner product, and operators) look different, but also that they may give rise to different physical predictions in their domain of applicability.

 In certain physical situations, such discrepancies are actually an asset of the polymeric theory rather than a drawback. In particular, the most prominent application of the polymer framework is in cosmology \cite{LivingReviewCosmo, AshtekarSingh,MathematicalLQC}, where the finite number of degrees of freedom for the metric of a homogeneous universe are quantized the loop way. Loop Quantum Cosmology predicts the same classical evolution far from the Big Bang regime but departs considerably from its counterpart Wheeler-DeWitt (Schr\"odinger) theory at the Planck scale. As an example, in several physical models it provides a mechanism to avoid the classical  Big Bang singularity and continue evolution through a \textquoteleft bounce' \cite{AshtekarCorichiSingh,APS}. Furthermore, the dynamics predicted is consistent with the inflation paradigm \cite{Agullo:2012}. Interestingly, in analogy with the full theory of LQG \cite{LOST}, the polymer representation has been singled out by symmetry arguments in some of these models \cite{AshtekarCampiglia2012}. These features are physically desirable and several tools are underway now to make LQC a falsifiable theory \cite{Agullo:2012, BojowaldEffectiveCosmology,  BojowaldCalcagni}.

However, in view of the overwhelming success of Schr\"odinger quantum mechanics in low energy physics, it may seem unjustified to apply the polymer quantization to non-relativistic mechanical systems. Indeed, one does not expect to extract any phenomenology from the corrections obtained by using the polymer framework in this context.
The polymer description comes equipped with a fundamental length  scale $\mu_0$ associated with a possible discreteness of space. In the gravitational context this is identified with the Planck length and (since no discreteness of space has been detected) in mechanical systems it must be much smaller than the natural length parameters associated to the system. Corrections are generically proportional to (powers of) $\mu_0$ and they become significant only in regimes where non-relativistic QM would be inapplicable \cite{Ashtekar}. 

Nevertheless, one may still apply the polymer description to simple mechanical systems for other reasons. Polymer quantum mechanics has been used as a toy model to illustrate and explore features of the full theory of LQG, like quantization ambiguities, its semiclassical limit \cite{Ashtekar, Corichi1}, its relation to the standard representations \cite{Corichi1,Corichi2,Fredenhagen} and symmetries \cite{Chiou2006}, and  to probe thermodynamical properties in this reduced context \cite{ChaconAcosta}.
Polymer quantum mechanical systems have also been studied to understand features of singular representations, as in \cite{HusainLoukoWinkler,KunstatterLoukoZiprick2008} and \cite{KunstatterLouko2012}, where inspired by cosmic singularity avoidance, they have been used to investigate effects on singular potentials and on boundary conditions respectively.
The polymer representation also illustrates the foundational relevance of inequivalent representations even in non-relativistic quantum mechanics, where in contrast to the Schr\"odinger description, Bohr's complementarity principle is fully implemented \cite{Halvorson}.
  
 One may  go a step further and assume the point of view where the general polymer or loop constructions constitute a more fundamental scheme for quantization, applicable to any physical system and not just gravity \cite{Thiemann1997,AshtekarPolymerScalar,Hossain,Lewandowski}. If this is the case for non-relativistic quantum mechanical systems, then all the well-tested constructions and results should follow from the polymer description in an appropriate limit.
  
In the present work we continue the study and exploration of polymer quantum mechanics along these lines. Inspired by \cite{Hossain}, we focus attention on one of the fundamental objects used to define dynamics in quantum systems: the propagator. The propagator encodes all the dynamical information of the system, its implementation in the polymer representation must fulfill properties analogous to the standard representation so that it gives consistent dynamical evolution, equivalent in a properly defined limit with Schr\"odinger unitary evolution. 

In this paper, we construct propagators and verify these consistency requirements  for the simplest quantum mechanical systems: a free particle on the real line and a particle in a one-dimensional box. These investigations were initiated in \cite{FloresProceedings}. For the benefit of the reader, before doing that,  in section \ref{PQM} we first give a self-contained review of the basic ingredients of the polymer representation and its dynamics. Section \ref{PPROPAGATORS} is the core of this work. There we revise the  consistency requirements propagators must satisfy and then,  we construct the propagators and verify explicitly these requirements for both systems. That we are able to give closed formulas for both propagators and verify explicitly their properties is a nontrivial result and a consistency check for the quantization scheme.  Finally, in section \ref{DISCUSSION}, we discuss our results and some implications and future work for polymer dynamics. 


\section{Polymer Quantum Mechanics} \label{PQM}


In order to understand the origin and exactly in what sense Polymer Quantum Mechanics differs from the textbook Schr\"odinger representation, we must recall Dirac's procedure for constructing a quantum theory out of a classical one.

The first step in constructing quantum mechanics from classical mechanics is given by the \textquoteleft quantization rule' that replaces the Poisson bracket of observables, i.e. functions on phase space, by the commutator of operators on an abstract Hilbert space $\mathcal{H}$. These operators are the corresponding observables of the quantum theory. For a particle with position $x$ and momentum $p$ this rule gives the canonical commutation relations (CCR's):
\begin{align}
[\hat{x},\hat{x}]=i\hbar\widehat{\{x,x\}}=\hat{0}\,,   &\qquad [\hat{p},\hat{p}]=i\hbar\widehat{\{p,p\}}=\hat{0}\,, \notag\\
[\hat{x},\hat{p}]=i\hbar&\widehat{\{x,p\}}=i\hbar \hat{1}\,.  \label{CCRs}
\end{align}

The next step in the quantization procedure is  to find a concrete Hilbert space $\mathcal{H}$ and a realization of $\hat{x}$ and $\hat{p}$ as hermitian operators acting on it, i.e.  to construct a representation of the CCR's. The standard choice for a point particle in one dimension is $\mathcal{H}_{Sch}=L^2(\mathbb{R},\md x)$, the set of complex-valued square integrable functions over $\mathbb{R}$, with the standard Lebesgue integral $\md x$, and with position and momentum operators $\hat{x}$ and $\hat{p}$ acting by multiplication and derivation respectively.

However, there are other choices for the Hilbert space $\mathcal{H}$, and/or the implementation of the position and momentum operators acting on it, and these other choices may or may not be (unitarily) equivalent to the Schr\"odinger representation.

To see how the polymer representation comes about, it is more convenient to work with the exponentiated versions  of $\hat{x}$ and $\hat{p}$, or more generally with the one-parameter families of unitary operators
\begin{equation} \label{generators}
  \widehat{U}_\mu=e^{i\mu\hat{x}}:=\sum_{n=0}^{\infty}\frac{1}{n!}(i\mu\hat{x})^n,  \qquad \qquad
\widehat{V}_\lambda=e^{i\lambda\hat{p}/\hbar}:=\sum_{n=0}^{\infty}\frac{1}{n!}(i\lambda\hat{p}/\hbar)^n,
\end{equation}
for arbitrary real parameters $\mu,\lambda\in\mathbb{R}$.
Unlike $\hat{x}$ and $\hat{p}$, if we define $\widehat{U}_\mu$ and $\widehat{V}_\lambda$ by their action on an arbitrary state or wave function $\psi\in\mathcal{H}_{Sch}$:
\begin{equation} \label{OperatorAction}
\widehat{U}_\mu\psi(x)=e^{i\mu x}\psi(x)\,, \qquad \qquad
\widehat{V}_\lambda\psi(x)=\psi(x+\lambda)\,,
\end{equation}
$\widehat{U}_\mu$ and $\widehat{V}_\lambda$ are well-defined operators on the whole Hilbert space $\mathcal{H}_{Sch}$, taking any (square integrable) wave function and sending it to another (square integrable) function\footnote{As opposed to e.g. $\hat{p}$ which can only act on differentiable functions and, like $\hat{x}$, may take a square integrable function and give another one whose integral diverges. From a mathematical standpoint,  it is more desirable to work with unitary (bounded) operators rather than only densely defined unbounded self-adjoint operators.}.

Acting on an arbitrary state it is easy to see that $\widehat{U}_\mu$ and $\widehat{V}_\lambda$ satisfy the product or composition rules:
\begin{align}
\widehat{U}_{\mu_1}\widehat{U}_{\mu_2}=\widehat{U}_{\mu_1+\mu_2}\,,  & \qquad  
\widehat{V}_{\lambda_1} \widehat{V}_{\lambda_2}=\widehat{V}_{\lambda_1+\lambda_2}\,, \notag \\
\widehat{U}_\mu\widehat{V}_\lambda=&e^{-i\mu\lambda}\widehat{V}_\lambda\widehat{U}_\mu\,,  \label{WEYL}
\end{align}
with the obvious properties (reality conditions): $\widehat{U}_\mu^\dagger=\widehat{U}_{-\mu}$ and  $\widehat{V}_\lambda^\dagger=\widehat{V}_{-\lambda}$.

In the Schr\"odinger representation, the latter products are equivalent to the canonical commutation relations (\ref{CCRs}), and define what mathematicians call a Weyl algebra.

Now one may step back and take (\ref{WEYL}) instead of (\ref{CCRs}) as the fundamental canonical commutation relations and look for Hilbert spaces and unitary operators thereon which satisfy (\ref{WEYL}).

Does it matter physically which representation of (\ref{WEYL}) we choose?. The celebrated Stone-von Neumann theorem \cite{ReedSimon11980} states that under certain regularity and irreducibility conditions on the operators and the representation, the answer is in the negative. Any two such representations are unitarily equivalent and therefore they produce the same physics. This is the reason why it is \textquoteleft sufficient' to work with the Schr\"odinger representation!.

However, motivated by some basic physical principle (as in \cite{Halvorson}), or inspired by a more fundamental theory (as in \cite{Ashtekar}), one may relax one or some of the regularity assumptions of the uniqueness theorem, or come up with a Hilbert space and operators that violate such conditions, therefore rendering a representation inequivalent to Schr\"odinger's.  This is precisely what is done in the polymer description.   

The central difference between Schr\"odinger and polymer quantization is the choice of a non-separable Hilbert space $\mathcal{H}_{poly}$. A Hilbert space with an uncountable orthonormal basis characterized by abstract kets $\left|\mu\right\rangle$ labelled by real numbers $\mu$, such that
\begin{equation} \label{PolymerInnerProduct}
\left\langle\mu|\nu\right\rangle=\delta_{\mu,\nu}\,,
\end{equation}
where as usual $\left\langle\mu|\nu\right\rangle$ denotes the inner product of kets $|\mu\rangle$ and $|\nu\rangle$, and $\delta_{\mu,\nu}$ is the Kronecker delta (not the Dirac distribution). 
An abstract vector  $\Psi$ element of $\mathcal{H}_{poly}$ is then expanded as $\sum_{i=1}^{\infty}\langle\mu_i|\Psi\rangle|\mu_i\rangle$, for some countable subset of basis kets $|\mu_i\rangle$.
 The basic operators $\widehat{U}_{\nu}$ and $\widehat{V}_\lambda$, implementing the canonical commutation relations (\ref{WEYL}), act on the basis vectors as:  
 \begin{equation} \label{operatorActions}
 \widehat{U}_{\nu}|\mu\rangle=e^{i\nu\mu}|\mu\rangle,  \qquad \widehat{V}_\lambda\left|\mu\right\rangle=\left|\mu-\lambda\right\rangle,
 \end{equation}
in analogy with (\ref{OperatorAction}). Their action on arbitrary vectors on $\mathcal{H}_{poly}$  is extended by linearity.
One can check by direct calculation that $\widehat{U}_\mu$ and $\widehat{V}_\lambda$ are well-defined unitary operators on $\mathcal{H}_{poly}$ and that indeed they satisfy (\ref{WEYL}). The position  operator $\hat{x}$, defined by its action on basis kets as $\hat{x}\left|\mu\right\rangle=\mu\left|\mu\right\rangle$ and extended by linearity, is such that $\widehat{U}_\mu=e^{i\mu\hat{x}}$ just as in the Schr\"odinger case. However, one of the key differences between the Schr\"odinger and polymer representations is that on the polymer Hilbert space there does not exist a hermitan operator $\hat{p}$ such that (\ref{generators}) is satisfied: the momentum operator $\hat{p}$ is not defined on $\mathcal{H}_{poly}$!.\footnote{Technically this is a consequence of the non-regularity of the polymer representation: the operator $\widehat{V}_\lambda$ is not weakly continuous with respect to $\lambda$, i.e. the matrix elements $\langle\mu|\widehat{V}_\lambda|\nu\rangle$ as functions of the real parameter $\lambda$ are not continuous: $\langle\mu|\widehat{V}_\lambda|\mu\rangle=\left\langle\mu|\mu-\lambda\right\rangle$ is $1$ if $\lambda=0$ and $0$ otherwise, so $\lim_{\lambda\to 0}\langle\mu|\widehat{V}_\lambda|\mu\rangle\neq \langle\mu|\widehat{V}_{\lambda=0}|\mu\rangle$. This precludes the existence of a hermitian generator for the $\widehat{V}_\lambda$'s \cite{Stone1932}. 

One may also exchange the definitions of $\widehat{U}_{\nu}$ and $\widehat{V}_\lambda$ in (\ref{operatorActions}) and obtain an inequivalent representation where the momentum operator is well-defined but the position operator is not \cite{Corichi1}.}

\begin{figure} 
      \begin{center}
      \includegraphics[width=8cm]{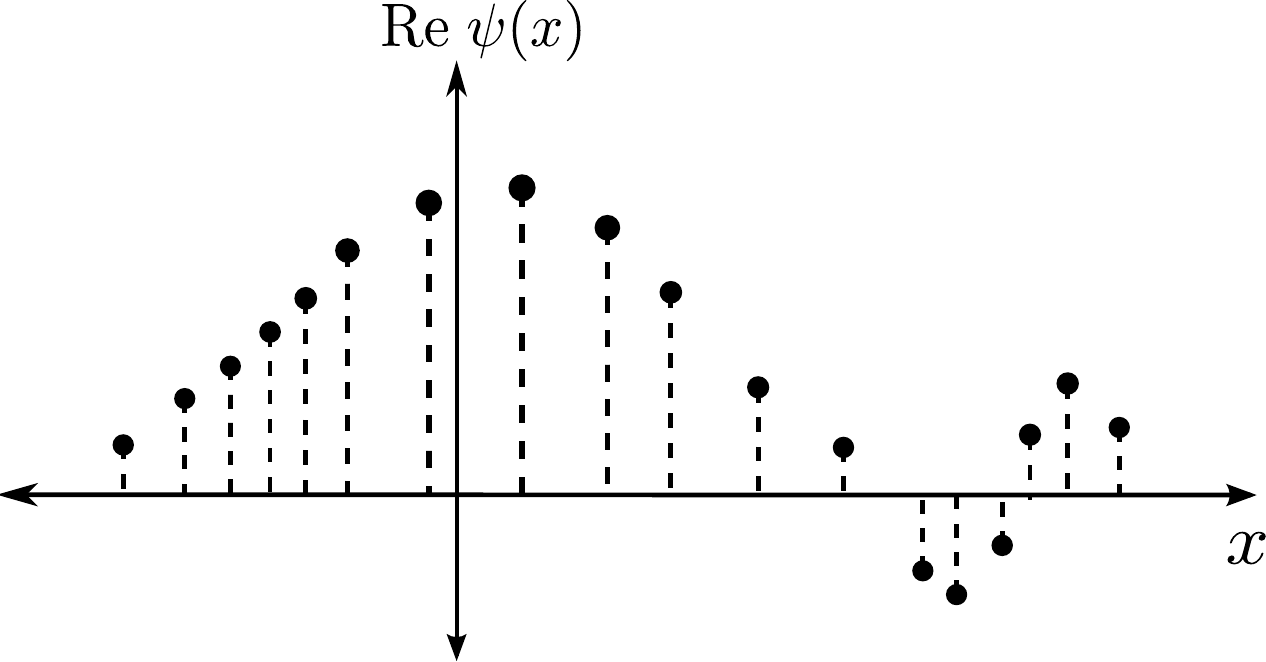}
      \caption{In the position representation, a polymer wave function is a modulated \textquoteleft Kronecker comb': $\psi(x)$ is nonzero at most on a countable number of points (not necessarily equally spaced).}
      \label{polymerWaveFnFig}
      \end{center}
\end{figure}

To realize the previous somewhat abstract characterization of $\mathcal{H}_{poly}$ in a concrete function space, in the position representation of $\mathcal{H}_{poly}$, one substitutes integrations of functions on the real line by discrete sums, so the normalizable wave functions $\psi$ are (complex-valued) functions on $\mathbb{R}$ that vanish at all but a countable number of points $\{x_n\}_{n=1}^\infty$, and such that $\sum_{n=1}^\infty|\psi(x_n)|^2<\infty$.\footnote{Technically, one replaces the Lebesgue measure $\md x$ on $\mathbb{R}$ by the counting measure $\md\mu_\text{count}$, so that $\mathcal{H}_{poly}=l^2(\mathbb{R})=L^2(\mathbb{R},\md\mu_\text{count})$. (Sometimes, in order to stress the analogy with LQG, $\mathbb{R}$ is also endowed with the discrete topology, but this is irrelevant for measure theoretic purposes) } 
The (uncountable) basis is then
\begin{equation}
\phi_\mu(x):=\left\langle x|\mu\right\rangle=\delta_{\mu,x}\,,
\end{equation}
so that an arbitrary wave function in $\mathcal{H}_{poly}$ may be expanded as a train of weighted Kronecker delta functions: $\psi=\sum_{n=1}^\infty\psi(x_n)\phi_{x_n}$ (Figure \ref{polymerWaveFnFig}). (Notice that while we only need a countable number of $\phi_{\mu_n}$'s to construct a particular wave function $\psi$, we do need all the uncountably many $\phi_\mu$'s to express the totality of wave functions in $\mathcal{H}_{poly}$ as discrete sums of Kronecker deltas). 

Finally, the position operator acts by multiplication on the basis vectors: $\hat{x}\phi_\mu=\mu\phi_\mu$, and $\widehat{V}_\lambda\phi_\mu=\phi_{\mu-\lambda}$ so that
\begin{equation}
\hat{x}\psi=\sum_{n=1}^\infty x_n\psi(x_n)\phi_{x_n}\,, \quad \text{and} \quad \widehat{V}_\lambda\psi=\sum_{n=1}^\infty\psi(x_n)\phi_{x_n-\lambda}\,.
\end{equation}

The momentum representation for $\mathcal{H}_{poly}$ is more subtle. In contrast with the Schr\"odinger representation, where the Fourier transform gives essentially the same function space for both position and momentum representations of $\mathcal{H}_{Sch}$ (square integrable wave functions on $\mathbb{R}$), \textquoteleft momentum wave functions' in the polymer representation are characterized differently from the position wave functions.

In the momentum representation, the orthonomal basis of the position operator eigenkets may be identified with plane waves:
 \begin{equation}
 \tilde{\phi}_\mu(p):=\left\langle p|\mu\right\rangle=\exp(i\mu p/\hbar),
\end{equation}
with the displacement and position operators acting on them by multiplication $\widehat{V}_\lambda=\exp(-i\lambda p/\hbar)$ and derivation $\hat{x}=-i\hbar\frac{\partial}{\partial p}$ respectively. 

The set of finite linear combinations of $\tilde{\phi}_\mu$'s is called the space of almost-periodic functions \cite{Corduneanu1989}, so in the momentum representation, $\mathcal{H}_{poly}$ is the completion of the space of almost-periodic functions with respect to the polymer inner product (\ref{PolymerInnerProduct}), that is the space of wave functions $\tilde{\psi}=\sum_{n=1}^\infty\psi(x_n)\tilde{\phi}_{x_n}$ such that $\sum_{n=1}^\infty|\psi(x_n)|^2<\infty$.

Notice that in contrast with the Schr\"odinger representation, position eigenkets are always normalizable in the polymer description. To reproduce the inner product (\ref{PolymerInnerProduct}) then, we may not use the standard Lebesgue integral $\md p$, but the modified formula\footnote{Using the theory of $C^*$-algebras and a theorem by Gel'fand, one can identify the completion space of almost periodic functions with the set of (continuous)  functions on a much bigger space $\bar{\mathbb{R}}_{Bohr}$. This space, called the Bohr compactification of the real line, contains the real numbers and is equipped with a binary operation that extends the addition of ordinary real numbers. As a (compact) group, $\bar{\mathbb{R}}_{Bohr}$ is also equipped with a natural notion of integration (the Haar measure $\md\mu_{Haar}$), so $\mathcal{H}_{poly}=L^2(\bar{\mathbb{R}}_{Bohr},\md\mu_{Haar})$. This is the preferred characterization of $\mathcal{H}_{poly}$ in the LQC literature, since it resembles more closely the constructions in LQG \cite{Velhinho}.}:
\begin{equation} \label{pInnerProduct}
\left\langle\mu|\nu\right\rangle=(\tilde{\phi}_\mu,\tilde{\phi}_\nu)=\lim_{L\to\infty}\frac{1}{2L}\int_{-L}^{L}\md p\,\tilde{\phi}_\mu^*(p)\tilde{\phi}_\nu(p).
\end{equation}


\subsection{Dynamics}

Given that momentum $\hat{p}$ is not a well-defined operator on $\mathcal{H}_{poly}$, to introduce dynamics one needs to construct operators $\widehat{p}_{\mu_o}$ (or $\widehat{p_{\mu_o}^2}$) that play the role of $\hat{p}$ (or $\widehat{p^2}$)  in the Hamiltonian for a particle of mass $m$ in a potential $\mathcal{V}(x)$:
\begin{equation}
H=\frac{p^2}{2m}+\mathcal{V}(x)\,.
\end{equation}

Inspired by the techniques used in lattice gauge theories and Loop Quantum Gravity, one fixes a length scale $\mu_0$ and approximates the operator $\widehat{p^2}$ using the classical expression
\begin{equation}
e^{i\mu_0 p/\hbar}+e^{-i\mu_0 p/\hbar}\approx 2-\mu_0^2p^2/\hbar^2\,,   \qquad \text{  for } p<<\hbar/\mu_0.
\end{equation}
The  exponentials may be directly quantized to $\widehat{V}_{-\mu_0}$ and $\widehat{V}_{\mu_0}$, so one defines
\begin{equation}
\widehat{p_{\mu_o}^2}:=\frac{\hbar^2}{\mu_0^2}\left[2-\widehat{V}_{\mu_0}-\widehat{V}_{-\mu_0}\right]\,.
\end{equation}
The limit $\lim_{\mu_0\to 0}\widehat{p_{\mu_o}^2}$ does not exist in $\mathcal{H}_{poly}$, so we really  cannot remove the  parameter $\mu_0$ to get a unique well-defined operator. $\mu_0$ becomes a free parameter of the theory.\footnote{From considerations in LQG, one may associate $\mu_0$ with a possible discreteness of space-time or alternatively, as in \cite{Corichi2}, consider it as a renormalization parameter.}  

We may now proceed as in the Schr\"odinger case, with dynamics determined by the Schr\"odinger equation:
\begin{equation}
i\hbar\frac{\partial \Psi}{\partial t}=\widehat{H}_{\mu_0}\Psi\,,
\end{equation}
whose stationary solutions $\Psi=e^{-iEt/\hbar}\psi$ are constructed from the energy eigenstates of the Hamiltonian operator \cite{Ashtekar}:
\begin{equation}
\widehat{H}_{\mu_0}:=\frac{\hbar^{2}}{2m\mu^{2}_0}\left[2-\widehat{V}_{\mu_0}-\widehat{V}_{-\mu_0}\right]+\mathcal{V}(\hat{x}).\label{PolymerHamiltonian}
\end{equation} 

The energy eigenvalue equation $\widehat{H}_{\mu_0}\psi=E\psi$ becomes a difference equation for the wave function $\psi(x)$ in the position representation:
\begin{equation}\label{DifferenceEq}
\psi(x+\mu_0)+\psi(x-\mu_0)=\left(2-\frac{2m\mu^{2}_0}{\hbar^2}\left(E-\mathcal{V}(x)\right)\right)\psi(x).
\end{equation}
In contrast, in the momentum representation, it is generically (at least for polynomial $\mathcal{V}(\hat{x})$) a differential equation for $\tilde{\psi}(p)$:
\begin{equation} \label{MomentumEigenvalueEq}
\left[\mathcal{V}\left(-i\hbar\frac{\partial}{\partial p}\right)+\frac{\hbar^2}{m
\mu^{2}_{0}}\left(1-\cos\left(\frac{\mu_{0}p}{\hbar}\right)\right)-
E\right]\tilde{\psi}(p)=0\,.
\end{equation}

Equation (\ref{DifferenceEq}) suggests a general solution may be constructed from simpler solutions taking nonzero values only at  regularly spaced points $x_n=x_0+n\mu_0$, for some $x_0\in\mathbb{R}$ on the real line. Indeed, given the values of $E$, $\psi(x_0)$ and $\psi(x_0+\mu_0)$ (or $\psi(x_0-\mu_0)$), we can use (\ref{DifferenceEq}) to construct a unique solution $\psi_{x_0}$ on the lattice $\gamma_{x_0,\mu_0}:=\{x_n\in\mathbb{R}|x_n=x_0+n\mu_0,\,n\in\mathbb{Z}\}$. A general solution of (\ref{DifferenceEq}) will then be a linear superposition of such solutions $\psi_{x_0}$ for $x_0\in[0,\mu_0)$.

\begin{figure} 
      \begin{center}
      \includegraphics[width=8cm]{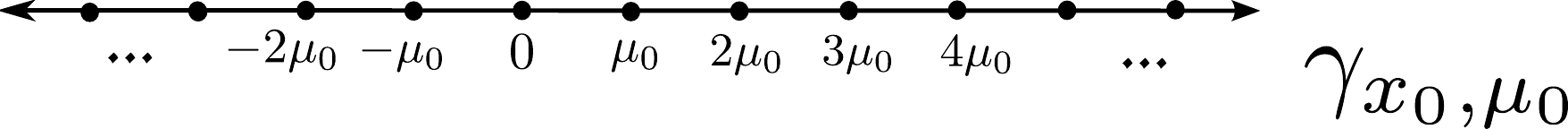}
      \caption{Dynamics is restricted to a regular lattice $\gamma_{x_0,\mu_0}$, with separation length $\mu_0$, and for simplicity here taken to include the origin $x_0=0$.}
      \label{latticeFig}
      \end{center}
\end{figure}

So fixing the length scale $\mu_0$, the form of the Hamiltonian (\ref{PolymerHamiltonian}) effectively allows us to restrict dynamics to a lattice $\gamma_{x_0,\mu_0}$ and work on a separable Hilbert space $\mathcal{H}_{\gamma_{x_0,\mu_0}}$ consisting of wave functions $\psi$ which are nonzero only on the lattice and such that $\sum_{n=-\infty}^{\infty}|\psi(x_n)|^2<\infty$ (Figure \ref{latticeFig}). This space has the countable basis $\{|x_n\rangle\}_{n\in\mathbb{Z}}$ \footnote{Spaces $\mathcal{H}_{\gamma_{x_0,\mu_0}}$ and $\mathcal{H}_{\gamma_{x_0',\mu_0}}$  are orthogonal for $x_0\neq x_0'$, and the full polymer Hilbert space may be decomposed as a direct sum of (separable) Hilbert spaces which are superselected by $\hat{H}_{poly}$:  
\[\mathcal{H}_{poly}=\bigoplus_{x_0\in[0,\mu_0)}\mathcal{H}_{\gamma_{x_0,\mu_0}}\]}
.

Strictly, the polymer Hamiltonian operator (\ref{PolymerHamiltonian}) is defined on all of $\mathcal{H}_{poly}$ and the most general solution to (\ref{DifferenceEq}) does not belong to a single $\mathcal{H}_{\gamma_{x_0,\mu_0}}$. However, if one wants to assign $\widehat{H}_{\mu_0}$ a consistent physical interpretation, one must restrict its action to a single $\mathcal{H}_{\gamma_{x_0,\mu_0}}$ and work on a fixed lattice $\gamma_{x_0,\mu_0}$ \cite{Corichi1}. This is what is generally done in the literature  (taking $x_0=0$ for simplicity) and what we will do here.

How does this restriction translate to the momentum representation? If we apply the unitary transformation: $\delta_{\mu,x}\mapsto\exp(i\mu p/\hbar)$, from  position to momentum representation, we obtain\footnote{In harmonic analysis or digital signal processing, this tranformation restricted to a fixed regular lattice is referred to as the discrete-time Fourier transform, or a special case of the $z$-tranform \cite{DSP}.}:
\begin{equation}
\psi(x)=\sum_{n=-\infty}^{\infty}\psi(x_n)\delta_{x_n,x} \,\mapsto\, \tilde{\psi}(p)=\sum_{n=-\infty}^{\infty}\psi(x_n)e^{ix_np/\hbar}\,.
\end{equation}
Working on $\gamma_{x_0,\mu_0}$ restricts momentum wave functions $\tilde{\psi}(p)$ to periodic functions of period $2\pi\hbar/\mu_0$ with the inner product formula (\ref{pInnerProduct}) reducing to
\begin{equation}
\left\langle x_n|x_l\right\rangle=(\tilde{\phi}_{x_n},\tilde{\phi}_{x_l})=\frac{\mu_0}{2\pi\hbar}\int_{-\pi\hbar/\mu_0}^{\pi\hbar/\mu_0}\md p\,\tilde{\phi}_{x_n}^*(p)\tilde{\phi}_{x_l}(p).
\end{equation}
and $p\in(-\pi\hbar/\mu_0,\pi\hbar/\mu_0)$, so effectively we are working with square integrable functions on the circle.


\section{Propagators in Polymer Quantum Mechanics} \label{PPROPAGATORS}


We now turn to the study of quantum propagators.
We define the propagator in the usual way \cite{Sakurai}. For time-independent Hamiltonians this is
\begin{equation}  \label{ThePropagator}
k(x_j,t;x_r,t_0):=\langle x_j,t|x_r,t_0\rangle=\left\langle
x_j\right|\exp\left\{-\frac{i}{\hbar}\widehat{H}_{\mu_0}(t-t_{0})\right\}\left|x_r\right\rangle,
\end{equation}
where here and in the following we work on a fixed lattice with $x_0=0$, so that $x_n=n\mu_0$. As already mentioned above,  based on the requirement of having a consistent physical interpretation for (\ref{PolymerHamiltonian}), we restrict $\widehat{H}_{\mu_0}$ to $\mathcal{H}_{x_0,\mu_0}$. We postpone the analysis of the unrestricted propagator on arbitrary lattices for later work.

Since we are now working on a separable Hilbert space $\mathcal{H}_{x_0,\mu_0}$, standard constructions and properties follow with just minor modifications. The propagator can be solved once we expand the initial position ket $|x_r\rangle$ in terms of the (complete) energy eigenkets $|E\rangle$ (or the eigenkets of an observable that commutes with $\widehat{H}_{\mu_0}$):
\begin{equation} \label{SpectralPropagator}
k(x_j,t;x_r,t_0)=\sum_E\langle x_j|E\rangle\langle E|x_r\rangle\exp\left\{-\frac{i}{\hbar}E(t-t_{0})\right\}\,.
\end{equation}
And it gives dynamical quantum evolution through the formula:
\begin{equation} \label{PolymerEvolution}
\psi(x_j,t)=\sum^{\infty}_{r=-\infty}k(x_j,t;x_r,t_{0})\psi(x_r,t_{0})\,.
\end{equation}

If the propagator is to implement well-defined quantum evolution, it must satisfy additional consistency requirements which follow directly from its definition:
\begin{enumerate}[i)]
\item From the completeness of energy eigenkets in (\ref{SpectralPropagator}), and (\ref{PolymerInnerProduct}) we must have 
\begin{equation} \label{Identity}
\lim_{t \rightarrow t_0}k(x_j,t;x_r,t_0)=\delta_{j,r}\,.
\end{equation}
\item Inserting the completeness relation $\sum_{n=-\infty}^\infty|x_n,t_1\rangle\langle x_n,t_1|=\hat{1}$ for position eigenkets at a fixed time $t_1$ in (\ref{ThePropagator}), we obtain the composition rule for $t_0<t_1<t$:
\begin{equation} \label{Composition}
 k(x_j,t;x_r,t_0)=\sum^{\infty}_{n=-\infty}k(x_j,t;x_n,t_1)k(x_n,t_1;x_r,t_0)\,.
 \end{equation}
\item $K(x_j,t;x_r,t_0):=\Theta(t-t_0)k(x_j,t;x_r,t_0)$, with $\Theta$ the Heaviside step function, is a Green's function for the Schr\"odinger operator $i\hbar\frac{\partial}{\partial t} - \widehat{H}_{\mu_0}$:
\begin{equation} \label{GreensFn}
\left(i\hbar\frac{\partial}{\partial t} - \widehat{H}_{\mu_0}\right)K(x_j,t;x_r,t_0)=i\hbar\delta(t-t_0)\delta_{j,r}\,.
\end{equation}
\end{enumerate}

In the following we will construct the propagator for the two simplest systems in quantum mechanics: a free particle and a particle in a box, and we will verify explicitly that such conditions are indeed satisfied.


\subsection{Propagator for the Free Particle}

To write down the propagator using the spectral formula (\ref{SpectralPropagator}), we must find the eigenvalues and eigenvectors of $\widehat{H}_{\mu_0}$ first. For a free particle $\mathcal{V}(x)=0$, so the energy eigenvalue equation (\ref{DifferenceEq}) reads
\begin{equation}
\psi(x+\mu_0)+\psi(x-\mu_0)=\left(2-\frac{2m\mu^{2}_0}{\hbar^2}E\right)\psi(x)\,,
\end{equation}
or defining $\mathcal{E}:=1-\frac{m\mu^{2}_0}{\hbar^2}E$ and using the short hand notation $\psi_n:=\psi(n\mu_0)$ for a wave function $\psi$ on a lattice $\gamma_{0,\mu_0}$:
\begin{equation} \label{LDE}
\psi_{n+1}-2\mathcal{E}\psi_n+\psi_{n-1}=0\,.
\end{equation}
This is a homogeneous second order linear difference equation with constant coefficients which may be solved by standard methods \cite{Elaydi}. Assuming a power form solution $\psi_n=\Lambda^n$ for some nonzero constant $\Lambda$ gives the characteristic equation
\begin{equation}
\Lambda^2-2\mathcal{E}\Lambda+1=0\,,
\end{equation}
with roots $\Lambda_{\pm}=\mathcal{E}\pm\sqrt{\mathcal{E}^2-1}$ . If $\Lambda_+\neq\Lambda_-$ then the general solution of (\ref{LDE}) is the linear combination
\begin{equation}
\psi_n=A\Lambda_+^n+B\Lambda_-^n\,,
\end{equation}
with arbitrary constants $A,B\in\mathbb{C}$. For equal roots $\Lambda=\Lambda_+=\Lambda_-$ the solution is
\begin{equation}
\psi_n=\Lambda^n(A+Bn)\,.
\end{equation}
There are four cases to analyze:  $\mathcal{E}^2>1$, $\mathcal{E}^2<1$, and $\mathcal{E}=\pm1$. Since $\widehat{H}_{\mu_0}$ and $\widehat{V}_{\mu_0}$ are commuting operators, they must have a common set of eigenvectors $\psi^E$. In particular $\psi^E$ must be such that they give  eigenvalues of unit norm for $\widehat{V}_{\mu_0}$.  Using this fact one may discard all but the complex root solutions
\begin{equation}
\psi_n=A\left(\mathcal{E}+i\sqrt{1-\mathcal{E}^2\,}\right)^n+B\left(\mathcal{E}-i\sqrt{1-\mathcal{E}^2\,}\right)^n\,,
\end{equation}
with
\begin{equation}
0\leq E \leq \frac{2\hbar^2}{m\mu^{2}_0}\,.
\end{equation}
Written in polar coordinates
\begin{align}
\psi_n^E=&\;A\left(e^{i\arccos(\mathcal{E})}\right)^n\;+\;B\left(e^{-i\arccos(\mathcal{E})}\right)^n \notag\\
=&\;Ae^{i n \arccos\left( 1-\frac{m\mu^{2}_0}{\hbar^2}E \right)}\;+\;Be^{-i n \arccos\left(1-\frac{m\mu^{2}_0}{\hbar^2}E\right)}\,. \label{GenXsolution}
\end{align}

On the other hand, the energy eigenvalue equation (\ref{MomentumEigenvalueEq}) in the momentum representation is algebraic and takes the form
\begin{equation}
\left[\frac{\hbar^2}{m
\mu^{2}_{0}}\left(1-\cos\left(\frac{\mu_{0}p}{\hbar}\right)\right)-
E\right]\tilde{\psi}(p)=0,
\end{equation}
giving the dispersion relations:
\begin{equation} \label{dispersion}
E=\frac{\hbar^2}{m\mu^{2}_{0}}\left(1-\cos\left(\frac{\mu_{0}p}{\hbar}\right)\right)\,, \qquad \text{or} \qquad
p_E=\frac{\hbar}{\mu_0}\arccos\left( 1-\frac{m\mu^{2}_0}{\hbar^2}E \right)\,,
\end{equation}
with $p_E\in(-\frac{\pi\hbar}{\mu_{0}},\frac{\pi\hbar}{\mu_{0}})$.
So from (\ref{GenXsolution}), the energy eigenfunctions in configuration and momentum space are respectively \cite{Corichi1}:
\begin{eqnarray}
\psi^E(x_n)=\exp\left\{\frac{i}{\hbar}x_n  p_E\right\}, &&
\tilde{\psi}^E(p)=\frac{2\pi\hbar}{\mu_{0}}\delta(p-p_E)\,.
\end{eqnarray}
There is, as expected, a two-fold degeneracy on the (continuous) energy spectrum, labeled as in the Schr\"odinger case by momentum $\pm |p_E|$. As emphasized before, solutions $\langle x|p_E\rangle:=\sum_ne^{ix_np_E/\hbar}\delta_{x_n,x}$
are eigenfunctions of the displacement operator on the lattice: $\widehat{V}_{\mu_0}|p_E\rangle=e^{i\mu_0p_E/\hbar}|p_E\rangle$, and we have the completeness relation  (in $\mathcal{H}_{\gamma_{x_0,\mu_0}}$) for the eigenkets of the Hamiltonian\footnote{Just as in the Schr\"odinger case, energy eigenfunctions for the free particle are not normalizable. $\langle x|p_E\rangle$ does not belong to $\mathcal{H}_{\gamma_{x_0,\mu_0}}$ and one has to use a Gel'fand triplet or rigged space construction \cite{Bohm} to define $|p_E\rangle$ rigorously.}:
\begin{equation}
\frac{\mu_{0}}{2\pi\hbar}\int^{\frac{\pi\hbar}{\mu_{0}}}_{-\frac{\pi\hbar}{\mu_{0}}}\md p_E\left|p_E\right\rangle\left\langle
p_E\right|=\hat{1}.\label{Ecompleteness}
\end{equation}

Using the completeness relation (\ref{Ecompleteness}) we may now evaluate the propagator (\ref{SpectralPropagator}) for the free particle:
\begin{align}
k&(x_j,t;x_r,t_{0})= \notag\\
&=\frac{\mu_{0}}{2\pi\hbar}\exp\left\{-\frac{i\hbar
(t-t_{0})}{m \mu^{2}_{0}}\right\}
\int^{\frac{\pi\hbar}{\mu_{0}}}_{-\frac{\pi\hbar}{\mu_{0}}}\md p_E\exp\left\{i\left[\frac{
\mu_{0}p_E}{\hbar}(r-j)+ \frac{\hbar (t-t_{0})}{m
\mu^{2}_{0}}\cos\left(\frac{\mu_{0}p_E}{\hbar}\right)
\right]\right\}\,.  
\end{align}

Defining for simplicity $\zeta=\pi+\mu_{0} p_E/\hbar$, and $z:=\frac{\hbar(t-t_{0})}{m \mu^{2}_{0}}$,
we recognize the integral form for Bessel functions $J_n(z)$ of integer order $n$ \cite{Arfken}:
\begin{align} 
k(x_j,t;x_r,t_{0})&=\frac{(-1)^{r-j}}{2\pi}\,\exp\{-iz\}
\int^{2\pi}_{0}\md\zeta \, \exp\left\{i\left((r-j)\zeta -z\cos
\zeta\right)\right\}\notag \\
&= i^{r-j}J_{r-j}(z)e^{-iz} 
\end{align}
or written explicitly
\begin{equation}
k(x_j,t;x_r,t_{0})=i^{r-j}J_{r-j}\left(\frac{\hbar (t-t_{0})}{m
\mu^{2}_{0}}\right)\exp\left\{-i\left(\frac{\hbar (t-t_{0})}{m
\mu^{2}_{0}}\right)\right\}.   \label{FreePropagator}
\end{equation}
As expected, this is symmetric with respect to $j$ and $r$ since $J_{-n}(z)=(-1)^nJ_n(z)$.

We may readily verify that using formula (\ref{PolymerEvolution}), the propagator (\ref{FreePropagator}) gives the correct time evolution for arbitrary states. This is verified explicitly for energy eigenfunctions  $\psi(x_{r},t_{0})=\psi^E(x_r)=\exp\left\{\frac{i}{\hbar}x_r p_E\right\}$:
\begin{align}
\psi(x_j,t)&=\sum^{\infty}_{r=-\infty}i^{r-j}J_{r-j}(z)\exp\{-iz\}\exp\left\{\frac{i}{\hbar}x_r
p_E\right\}  \notag\\
&=\exp\{-iz\}\exp\left\{\frac{i}{\hbar}x_j
p_E\right\}\sum^{\infty}_{l=-\infty}i^{l}J_{l}(z)\exp\left\{\frac{i l \mu_{0}
p_E}{\hbar}\right\} \notag\\
&=\exp\{-iz\}\exp\left\{\frac{i}{\hbar}x_j
p_E\right\}\exp\left\{iz\,\cos\left(\frac{\mu_{0}p_E}{\hbar}\right)\right\} \notag\\
&=\exp\left\{-\frac{i}{\hbar}E(t-t_{0})\right\}\psi^E(x_{j}),\label{ec:5}
\end{align}
where we have substituted $r=l+j$, and used the Jacobi-Anger expansion \cite{Arfken}:
\begin{equation} \label{Jacobi-Anger}
e^{iz \cos \varphi}=\sum_{n=-\infty}^\infty i^nJ_n(z)e^{in\varphi}\,.
\end{equation}


\subsubsection{Properties}

Next we check explicitly that consistency conditions (\ref{Composition}-\ref{GreensFn}) are indeed satisfied for the free particle polymer propagator (\ref{FreePropagator}). They all follow from properties of Bessel functions \cite{Arfken,Watson}.

For (\ref{Identity}) we have
\begin{eqnarray}
\lim_{t \longrightarrow t_0}k(x_j,t;x_r,t_0)&=&\lim_{t
\longrightarrow t_0}i^{r-j}J_{r-j}\left(\frac{\hbar (t-t_{0})}{m
\mu^{2}_{0}}\right)\exp\left\{-i\left(\frac{\hbar (t-t_{0})}{m
\mu^{2}_{0}}\right)\right\}\nonumber\\
&=&i^{r-j}J_{r-j}(0).\label{ec:6}
\end{eqnarray}
and we know  $J_n(0)=1$, for $n=0$ and $J_n(0)=0$ for $n\neq 0$. So indeed the right hand side of  (\ref{ec:6}) is one for $r=j$, and cero if $r\neq j$.

The composition rule (\ref{Composition}) follows directly from the identities 
\begin{equation}
J_m(u+v)=\sum^{\infty}_{q=\infty}J_q(u)J_{m-q}(v).
\end{equation}

For condition (\ref{GreensFn}) we may start from the expression $\left\langle x_j\right|\left(i\hbar\partial/\partial t -\widehat{H}_{\mu_0}\right)\left|\psi(t)\right\rangle$, with $\left|\psi(t)\right\rangle=\exp\left\{-i\widehat{H}_{\mu_0}(t-t_{0})/\hbar\right\}\left|x_r\right\rangle$. Substituting propagator $k(x_j,t;x_r,t_0)\to K(x_j,t;x_r,t_0):=\Theta(t-t_0)k(x_j,t;x_r,t_0)$ gives 
\begin{align}
i\hbar\frac{\partial K(x_j,t;x_r,t_0)}{\partial t}&-\frac{\hbar^{2}}{2m\mu_{0}^{2}}\left[2K(x_j,t;x_r,t_0) - 
K(x_{j+1},t;x_r,t_0) - K(x_{j-1},t;x_r,t_0)\right]=  \notag\\ \notag\\
=\,i\hbar\delta(t-t_0&)k(x_j,t;x_r,t_0) \notag\\ 
+ \hbar\Theta&(t-t_0)i^{r-j+1}e^{-iz}  
\left[\frac{d}{dt}J_{r-j}(z)- \frac{\hbar}{2m\mu^{2}_{0}} \left(J_{r-j-1}(z)-J_{r-j+1}(z)\right)\right]\,.  \notag
\end{align}
Using identities $J_{m-1}(u)-J_{m+1}(u)=2\frac{d}{du}J_{m}(u)$, and property (\ref{ec:6}), the right hand side reduces to 
\[
i\hbar\delta(t-t_0)\delta_{j,r}\,,
\]
so indeed, $K(x_j,t;x_r,t_0)$ is  the Green's function for the corresponding difference equation.


\subsubsection{Relation to the Schr\"odinger representation} \label{CONTINUUMLIMIT}
The length scale $\mu_0$ is a free parameter of the polymer representation and is sometimes associated with the Planck scale and its non-zero value considered a consequence of a possible discreteness of space. So one expects that  the polymer formulation should reduce to the Schr\"odinger representation in the \textquoteleft limit' $\mu_0 \to 0$ . However, this is a delicate issue, as already pointed out in \cite{Corichi2}, and one must be careful in taking this limit.

In the problem at hand, the limit  should be taken so that the separation $x_r-x_j=\mu_0(r-j)$ between initial and final points $x_r$ and $x_j$ in (\ref{FreePropagator}) is kept fixed. Thus,  along $\mu_0 \to 0$, we must take $l\to \infty$, where $l:=|r-j|$ is the number of points between $x_j$ and $x_r$, and the order or index for the Bessel function appearing in the propagator expression (\ref{FreePropagator}). Since $\mu_0^{-2}$ appears in the argument of these Bessel functions, the analysis requires in principle asymptotic expansions for Bessel functions with large indices and large arguments. 
Nevertheless, in the regime where the polymer description should approximate Schr\"odinger dynamics,
the argument grows faster than the order. So in the end, the asymptotic behavior of the Bessel functions for large values of the argument $z=\frac{\hbar (t-t_{0})}{m\mu^{2}_{0}}$ \cite{Watson}:  
\begin{eqnarray}
J_{l}(z)&\approx&\left(\frac{2}{\pi
z}\right)^{\frac{1}{2}}\Bigl[\cos\left(z \mp\frac{l\pi}{2}-\frac{\pi}{4}\right)
\sum^{\infty}_{n=0}\frac{(-1)^{n}}{(2n)!(2z)^{2n}}
\frac{\Gamma\left(l+2n+\frac{1}{2}\right)}
{\Gamma\left(l-2n+\frac{1}{2}\right)} \nonumber\\ &&
-\sin\left(z \mp\frac{l\pi}{2}-\frac{\pi}{4}\right)
\sum^{\infty}_{n=0}\frac{(-1)^{n}}{(2n+1)!(2z)^{2n+1}}
\frac{\Gamma\left(l+(2n+1)+\frac{1}{2}\right)}
{\Gamma\left(l-(2n+1)+\frac{1}{2}\right)}\Bigr],\label{ec:8}
\end{eqnarray}
determines the limit of the polymer propagator in (\ref{FreePropagator}).
One can expect this behavior from the relation
\begin{equation}
\frac{z}{l}=\frac{\lambda}{\mu_0}
\end{equation}
obtained by combining $p=m(x_j-x_r)/(t-t_0)$ and the de Broglie wavelength $\lambda=2\pi\hbar/p$. Considering for reference typical diffraction experiments of electrons and neutrons: $\lambda\sim 10^{-10}m$, and taking $\mu_0$ in the order of the planck length $\ell_P=1.610^{-35}m$, gives $z/l\sim 10^{25}$, so this asymptotic expansion is valid even if one takes $\mu_0$ several orders of magnitude above the Planck scale. 

Now,  if we  define 
\begin{align}
\beta:=z \mp&\frac{\pi(r-j)}{2}-\frac{\pi}{4}, \qquad  P_l(z):=\sum^{\infty}_{n=0}\frac{(-1)^{n}}{(2n)!(2z)^{2n}}
\frac{\Gamma\left(l+2n+\frac{1}{2}\right)}{\Gamma\left(l-2n+\frac{1}{2}\right)},  \notag\\
\text{ and } &\quad  Q_l(z):=\sum^{\infty}_{n=0}\frac{(-1)^{n}}{(2n+1)!(2z)^{2n+1}}
\frac{\Gamma\left(l+(2n+1)+\frac{1}{2}\right)}
{\Gamma\left(l-(2n+1)+\frac{1}{2}\right)}, 
\end{align}
we obtain for (\ref{FreePropagator}):
\begin{equation}
k(x_j,t;x_r,t_0)\approx\sqrt{\frac{m}{2i\pi\hbar(t-t_0)}}\mu_{0}
\left[(P_l(z)+iQ_l(z))+(P_l(z)-iQ_l(z))e^{-2i\beta}\right].
\end{equation}

Furthermore, inserting the completeness relation (\ref{Ecompleteness}) for the energy eigenvectors in the polymer inner product of position eigenkets (\ref{PolymerInnerProduct}), we see that 
\begin{equation}
\lim_{\mu_0\rightarrow 0} \frac{\left\langle x_j|x_r\right\rangle_{poly}}{\mu_0}=\left\langle
x_j|x_r\right\rangle_{Schr}.
\end{equation} 
So to take the correct limit of the propagator we must divide it by $\mu_0$.

$P_l(z)$ and $Q_l(z)$ can be expressed as
\begin{eqnarray}
P_l(z)&=&\sum^{\infty}_{n=0}\frac{(-1)^{n}}{(2n)!}\left(\frac{
l^{2}}{2z}\right)^{2n}\left(1-\frac{1}{4l^{2}}\right)\left(1-\frac{3^{2}}{4l^{2}}\right)...
\left(1-\frac{(4n-1)^{2}}{4l^{2}}\right),\label{ec:9}\\
Q_l(z)&=&\sum^{\infty}_{n=0}\frac{(-1)^{n}}{(2n+1)!}\left(\frac{l^{2}}{2z}\right)^{2n+1}
\left(1-\frac{1}{4l^{2}}\right)...
\left(1-\frac{(4n+1)^{2}}{4l^{2}}\right), \label{ec:10}
\end{eqnarray}
and
\begin{equation}\label{ec:11}
e^{-2i\beta}=i\frac{(-1)^{l}}{l}l\exp\left\{
-\frac{2i\hbar(t-t_0)}{m(x_j-x_r)^{2}}l^{2}\right\}.
\end{equation}
We must now consider both limits  $\mu_0\rightarrow 0$ and $l\rightarrow \infty$. Taking the latter in (\ref{ec:9}) and (\ref{ec:10}) we get
\begin{equation}
P_l(z)\pm iQ_l(z)\sim \exp\left\{\pm i\frac{l^2}{2z}\right\}. 
\end{equation}
One can further see that  in this limit the term proportional to $e^{-2i\beta}$ makes no contribution to the propagator:
 we have the (distributional) expression
\begin{equation}
e^{-2i\beta}\sim i\cdot 0\cdot \sqrt{\frac{\pi}{i}}\,\delta\left(\sqrt{\frac{2\hbar(t-t_0)}{m(x_j-x_r)^{2}}}\,\right)= 0
\end{equation}
from (\ref{ec:11}) and $\delta(X)=\lim_{l\to\infty}\sqrt{\frac{i}{\pi}}l\exp(-il^2X^2)$.

Combining these results, we obtain the usual Schr\"odinger expression for the propagator of a free particle:
\begin{equation}  \label{FreeSchrodingerPropagator}
\lim_{l\rightarrow\infty}\lim_{\mu_0\rightarrow 0}\frac{k(x_j,t;x_r,t_0)}{\mu_0}=\sqrt{\frac{m}{2i\pi\hbar(t-t_0)}}\exp\left\{\frac{im(x_j-x_r)^{2}} {2\hbar(t-t_0)}\right\}.
\end{equation}


\subsubsection{Propagator in momentum space}

To compute the propagator in momentum space we may proceed as before to directly evaluate the matrix elements from the expression
\begin{equation}\label{ec:12}
G(p,t;p^{\prime},t_0):=\left\langle
p\right|\exp\left\{-\frac{i}{\hbar}\widehat{H}_{\mu_0}(t-t_{0})\right\}\left|p^{\prime}\right\rangle,
\end{equation}
or do a \textquoteleft Fourier transform' of the corresponding formula in configuration space:
\begin{equation}\label{ec:13}
G(p,t;p^{\prime},t_0):=\sum_{j,r}\left\langle
p|x_j\right\rangle\left\langle
x_j\right|\exp\left\{-\frac{i}{\hbar}\widehat{H}_{\mu_0}(t-t_{0})\right\}\left|x_r\right\rangle\left\langle
x_r|p^{\prime}\right\rangle.
\end{equation} 
Both procedures give the same final form for the propagator:
\begin{equation}
G(p,t;p^{\prime},t_0)=\frac{2\pi\hbar}{\mu_0}e^{-\frac{i}{\hbar}E_{p^{\prime}}(t-t_0)}\delta(p^{\prime}-p).
\end{equation}

Using similar considerations as above, one may readily verify this expression gives consistent evolution and reduces to the Schr\"odinger propagator in the limit $\mu_0\rightarrow 0$.


\subsection{Particle in a Box}

We consider now a particle in a box or potential well of length $L=N\mu_0$:
\[
\mathcal{V}(x)=\left\{\begin{array}{ll}
                        0       & \mbox{if $0<x<N\mu_0$};\\
                        \infty  & \mbox{otherwise}.
                        \end{array} \right.
\] 
So as usual, we must impose boundary conditions $\psi(x_0)=\psi(x_N)=0$ on the general free particle solution (\ref{GenXsolution}). This gives $A=-B$ and the discrete set of energy eigenvalues \cite{ChaconAcosta}:
\begin{equation} \label{Ebox}
E_l=\frac{\hbar^2}{m\mu_0^2}\left(1-\cos\left(\frac{l\pi}{N}\right)\right), \qquad \mbox{ for $l=1,2,\dots,N-1$}.
\end{equation}
After normalization $||\psi^{E_l}||^2=\sum_{n=0}^{N}|\psi^{E_l}_n|^2=1$ the corresponding eigenstates are
\begin{equation} \label{psibox}
\psi^{E_l}(x_n)=\sqrt{\frac{2}{N}}\sin\left(\frac{l\pi x_n}{N\mu_0}\right).
\end{equation}
Notice that unlike the Schr\"odinger case, because of the discreteness, the energy is bounded and we only have a finite number of eigenstates.

Using the spectral formula (\ref{SpectralPropagator}) and (\ref{Ebox})-(\ref{psibox}), the propagator for a particle in a box is given by the expression 
\begin{equation}\label{BoxPropagator}
k_\text{Box}(x_j,t;x_r,t_0)=\frac{2}{N}\sum^{N-1}_{n=1}\sin\left(\frac{n\pi j}{N}\right)\sin\left(\frac{n\pi r}{N}\right)e^{-iz\left[1-\cos\left(\frac{n\pi}{N}\right)\right]},
\end{equation}
where again we have defined $z=\frac{\hbar(t-t_0)}{m\mu^{2}_0}$.

\subsubsection{Properties}

Once more, all consistency properties can be readily verified in this case. The limit $t\to t_0$ for the propagator is just the completeness relation for the energy eigenfunctions
\begin{equation}\label{eq:cicp}
\lim_{t\rightarrow t_0}k_\text{Box}(x_j,t;x_r,t_0)=\frac{2}{N}\sum^{N-1}_{n=1}\sin\left(\frac{n\pi j}{N}\right)\sin\left(\frac{n\pi r}{N}\right)=\delta_{j,r}\,.
\end{equation}

From this, we may check tacitly the propagator gives the correct time evolution for an energy eigenstate $\psi(x_r,t_0)=\psi^{E_s}(x_r)$. Equation (\ref{PolymerEvolution}) gives
\begin{align}
\psi(x_j,t)&=\sum^{N-1}_{r=1}k_\text{Box}(x_j,t;x_r,t_0)\psi(x_r,t_0) \nonumber\\
           &=\left(\frac{2}{N}\right)^{\frac{3}{2}}\sum^{N-1}_{r,n=1}\sin\left(\frac{n\pi j}{N}\right)\sin\left(\frac{n\pi r}{N}\right)\sin\left(\frac{s\pi r}{N}\right)\,e^{-iz\left[1-\cos\left(\frac{n\pi}{N}\right)\right]} \nonumber\\ 
           &=\left(\frac{2}{N}\right)^{\frac{3}{2}}\sum^{N-1}_{n=1}\sin\left(\frac{n\pi j}{N}\right)e^{-iz\left[1-\cos\left(\frac{n\pi}{N}\right)\right]} \sum^{N-1}_{r=1}\sin\left(\frac{n\pi r}{N}\right)\sin\left(\frac{s\pi r}{N}\right) \nonumber\\&= \left(\frac{2}{N}\right)^{\frac{3}{2}}\sum^{N-1}_{n=1}\sin\left(\frac{n\pi j}{N}\right)e^{-iz\left[1-\cos\left(\frac{n\pi}{N}\right)\right]}\frac{N}{2}\delta_{n,s} \nonumber \\           
           &= \sqrt{\frac{2}{N}}\sin\left(\frac{n\pi j}{N}\right)e^{-iz\left[1-\cos\left(\frac{n\pi}{N}\right)\right]}\,. \nonumber 
\end{align}

For the composition property (\ref{Composition}), again using the completeness relation (\ref{eq:cicp}), and defining $z=\frac{\hbar(t-t_1)}{m\mu^{2}_0}$ and $z_1=\frac{\hbar(t_1-t_0)}{m\mu^{2}_0}$, we have
\begin{align}
\sum^{N-1}_{s=1}k_\text{Box}(x_j,t;x_s,t_1&)k_\text{Box}(x_s,t_1;x_r,t_0)= \notag\\
 =\left(\frac{2}{N}\right)^2&\,\sum^{N-1}_{n=1}\sin\left(\frac{n\pi j}{N}\right)\exp\{-i(z+z_1)+iz\cos\left(\frac{n\pi}{N}\right)\}\times \notag\\ 
&\times\sum^{N-1}_{v=1}\sin\left(\frac{v\pi r}{N}\right)\exp\{iz_1\cos\left(\frac{v\pi}{N}\right)\}\sum^{N-1}_{s=1}\sin\left(\frac{v\pi s}{N}\right)\sin\left(\frac{n\pi s}{N}\right) \notag\\
=\frac{2}{N}\sum^{N-1}_{n=1}&\sin\left(\frac{n\pi j}{N}\right)\sin\left(\frac{n\pi r}{N}\right)e^{-i(z+z_1)\left[1-\cos\left(\frac{n\pi}{N}\right)\right]}  \notag\\ \notag\\
=k_\text{Box}(x&_j,t;x_r,t_0).\nonumber 
\end{align}

We can further check $K_\text{Box}:=\Theta(t-t_0)k_\text{Box}$  is the Green's function for the operator
 $i\hbar\partial/\partial t-\widehat{H}_{\mu_0}$, with boundary and initial conditions corresponding to the system of a particle in a box:
\[
i\hbar\frac{\partial K_\text{box}}{\partial t}=i\hbar\delta(t-t_0)k_\text{Box}+\frac{2\Theta(t-t_0)}{N}\sum^{N-1}_{n=1}E_n\sin\left(\frac{n\pi j}{N}\right)\sin\left(\frac{n\pi r}{N}\right)e^{-iz\left[1-\cos\left(\frac{n\pi}{N}\right)\right]},
\]
and
\begin{eqnarray}
\widehat{H}_{\mu_0}K_\text{Box} &=& \frac{\hbar^2}{m\mu^{2}_{0}}\frac{\Theta(t-t_0)}{N}\sum^{N-1}_{n=1}\sin\left(\frac{n\pi r}{N}\right)e^{-iz\left[1-\cos\left(\frac{n\pi}{N}\right)\right]}\times\nonumber\\
& & \times \left\{2\sin\left(\frac{n\pi j}{N}\right)-\sin\left[\frac{n\pi (j+1)}{N}\right]-\sin\left[\frac{n\pi (j-1)}{N}\right]\right\}\nonumber\\
&=&\frac{2\Theta(t-t_0)}{N}\sum^{N-1}_{n=1}E_n\sin\left(\frac{n\pi j}{N}\right)\sin\left(\frac{n\pi r}{N}\right)e^{-iz\left[1-\cos\left(\frac{n\pi}{N}\right)\right]},    \notag    
\end{eqnarray}
from which we obtain
\[
\left(i\hbar\frac{\partial}{\partial t}-\widehat{H}_{\mu_0}\right)K_\text{Box}=i\hbar\delta(t-t_0)k_\text{Box}.
\]
The right hand side is nonzero only when $t=t_0$, so from (\ref{eq:cicp}) we may replace it by $i\hbar\delta(t-t_0)\delta_{j,r}$. Furthermore,  (\ref{BoxPropagator}) satisfies the boundary conditions
\begin{eqnarray}\label{eq:cfcp1}
&k_\text{Box}(x_j,t;x_r = 0 ,t_0) = k_\text{Box}(x_j,t;x_r = N,t_0)=0\,,&\\ \label{eq:cfcp2}
& k_\text{Box}(x_j = 0,t;x_r,t_0) = k_\text{Box}(x_j = N,t;x_r,t_0)=0\,.&
\end{eqnarray} 
 So indeed $K_\text{Box}$ is a Green's function. 
 
 Now we proceed to analyze an interesting result that is known to hold in the Schr\"odinger case, namely that one can relate a series of free propagators with that of the particle in the box.


\subsubsection{Method of images}

As first pointed out by Pauli \cite{Pauli}, the propagator for a particle in a box (\ref{BoxPropagator}) may also be derived from the propagator of the free particle (\ref{FreePropagator}) using the method of images.

We first construct the propagator for a particle with periodic boundary conditions:
\[
\psi(x)=\psi(x+2N\mu_0)
\]
by adding  infinitely many images of the free particle to the left and right of the interval $-N\mu_0\leq x\leq N\mu_0$:
\begin{align}
K_\text{P}(x_j,t;x_r,t_0):=&\sum^{\infty}_{k=-\infty}K(x_j,t;x_r + 2kN\mu_0,t_0) \notag\\
=&\Theta(t-t_0)\,e^{-iz}\sum^{\infty}_{k=-\infty}i^{j-r-2kN}J_{j-r-2kN}(z). \label{eq:ppp}
\end{align}   
Indeed, this has period $2N\mu_0$ in $x_j$ and $x_r$, and  it satisfies the linear equation (\ref{GreensFn}). The propagator for a particle in a box is twice the odd part of the propagator for the periodic particle:
\begin{eqnarray}\label{BoxPropagatorIm}
k_\text{Box}(x_j,t;x_r,t_0) &=& k_\text{P}(x_j,t;x_r ,t_0)-k_\text{P}(x_j,t;-x_r ,t_0) \nonumber\\
&=& e^{-iz}\sum^{\infty}_{k=-\infty}\left\{i^{j-r-2kN}J_{j-r-2kN}(z)\right.
-\left.i^{j+r-2kN}J_{j+r-2kN}(z)\right\}.
\end{eqnarray} 


With a little extra work, we can make direct contact with formula (\ref{BoxPropagator}) using the Jacobi-Anger expansion (\ref{Jacobi-Anger}) and basic trigonometric identities. Starting from the spectral formula (\ref{BoxPropagator}):
\begin{align}
k_\text{box}(x_j,t;x_r,t_0)&=\frac{2}{N}\sum^{N-1}_{n=1}\sin\left(\frac{n\pi j}{N}\right)\sin\left(\frac{n\pi r}{N}\right)e^{-iz\left[1-\cos\left(\frac{n\pi}{N}\right)\right]} \notag\\
&=\frac{2}{N}e^{-iz}\sum^{N-1}_{n=1}\sum^{\infty}_{k=-\infty}i^{k}J_{k}(z)\sin\left(\frac{n\pi j}{N}\right)\sin\left(\frac{n\pi r}{N}\right)\cos\left(\frac{n\pi k}{N}\right) \notag\\
&=\frac{1}{N}e^{-iz}\sum^{N-1}_{n=1}\sum^{\infty}_{k=-\infty}i^{k}J_{k}(z)\left\{\cos\left[\frac{n\pi(j-r-k)}{N}\right]-\cos\left[\frac{n\pi (j+r-k)}{N}\right]\right\}.  \label{FromStoI}
\end{align}
We may now exchange the finite and infinite sums. The finite sums $\sum^{N-1}_{n=1}\cos\left(\frac{n\pi l}{N}\right)$ for integer $l\in\mathbb{Z}$, may be evaluated using the geometric sum $1+r+r^2+...+r^{N-1}=\frac{1-r^N}{1-r}$
 as follows 
\begin{align}
\sum^{N-1}_{n=1}\cos\left(\frac{n\pi l}{N}\right)&=\frac{1}{2}\left\{\sum^{N-1}_{n=1}e^{i\frac{n\pi l}{N}}+\sum^{N-1}_{n=1}e^{-i\frac{n\pi l}{N}}\right\} \notag\\
&=\frac{1}{2}\left\{\frac{1-(-1)^{l}}{1-e^{i\frac{\pi l}{N}}}+\frac{1-(-1)^{l}}{1-e^{-i\frac{\pi l}{N}}}\right\}-1 \notag\\
&=\frac{1}{2}\left[1-(-1)^{l}\right]-1, \notag
\end{align}
which is valid for $e^{\pm i\frac{\pi l}{N}}\neq 1$, that is for $l\neq 2sN$, for some integer $s\in\mathbb{Z}$.
Hence
\[
 \sum^{N-1}_{n=1}\cos\left(\frac{n\pi l}{N}\right) = -1+\left\{
        \begin{tabular}{cc} 
        	$\frac{1}{2}[1-(-1)^{l}]$,&for $ l\neq 2sN$, \\
        	$N$,&if $l=2sN$. \\
        \end{tabular} \right. 
\]
Since $\frac{1}{2}[1-(-1)^{l}]=0$ for $l=2sN$,  we may also write
\begin{equation}\label{eq:ri}
\sum^{N-1}_{n=1}\cos\left(\frac{n\pi l}{N}\right)=-1+\sum^{\infty}_{s=-\infty}\left\{N\delta_{l,2sN}+\frac{1}{2}[1-(-1)^{l}]\delta_{l,s}\right\}.
\end{equation}

Substituting (\ref{eq:ri}) in (\ref{FromStoI}) we get
\begin{align}
k_\text{Box}(x_j,t;x_r,t_0)=\frac{1}{N}\,e^{-iz}\sum^{\infty}_{k=-\infty}i^{k}J_{k}(z)& \sum^{\infty}_{s-\infty}\left\{N\delta_{j-r-k,2sN}+\frac{1}{2}[1-(-1)^{j-r-k}]\delta_{j-r-k,s} \right. \notag\\
&\left.-N\delta_{j+r-k,2sN}-\frac{1}{2}[1-(-1)^{j+r-k}]\delta_{j+r-k,s}\right\} \nonumber\\
=e^{-iz}\sum^{\infty}_{s=-\infty}\left\{i^{j-r-2sN}\right.&\left.J_{j-r-2sN}(z)-i^{j+r-2sN}J_{j+r-2sN}(z)\right\}+ \notag\\
+\frac{e^{-iz}}{2N}\sum^{\infty}_{s=-\infty}[1-(-1&)^{s}]\left\{i^{j-r-s}J_{j-r-s}(z)-i^{j+r-s}J_{j+r-s}(z)\right\}. \label{Spectral2Image}
\end{align}

Relabeling indices, the last sum is easily seen to vanish:
\begin{align}
\sum^{\infty}_{s=-\infty}&[1-(-1)^{s}]\left\{i^{j-r-s}J_{j-r-s}(z) -  i^{j+r-s}J_{j+r-s}(z)\right\}= \notag\\
&=\sum^{\infty}_{s=-\infty}[1-(-1)^{s}]\left\{i^{j-r-s}J_{j-r-s}(z)- i^{j+r+s}J_{j+r+s}(z)\right\}  \notag\\
&=\sum^{\infty}_{l=-\infty}[1-(-1)^{l+r}]\left\{i^{j-l}J_{j-l}(z) - i^{j+l}J_{j+l}(z)\right\}  \notag\\
&=\sum^{\infty}_{l=-\infty}[1-(-1)^{l+r}]\left\{i^{j-l}J_{j-l}(z)-  i^{j-l}J_{j-l}(z)\right\}=0\,,\notag
\end{align}
and we regain propagator formula (\ref{BoxPropagatorIm}) from (\ref{Spectral2Image}).

One  may also  verify  consistency properties  explicitly using this formula \cite{FloresThesis}.


\subsubsection{Relation to the propagator in the Schr\"odinger representation}

The relation between the polymer and the Schr\"odinger propagator has some subtleties in this case.
Using formula (\ref{BoxPropagatorIm}) and the \textquoteleft continumm limit' of the free particle propagator (\ref{FreeSchrodingerPropagator}) we may formally write
\begin{align}
\lim_{\substack{N\rightarrow\infty,\,\mu_0\rightarrow 0 \\ N\mu_0=L}}\frac{k_\text{Box}(x_j,t;x_r,t_0)}{\mu_0}=\sqrt{\frac{m}{2i\pi\hbar(t-t_0)}}\sum_{k=-\infty}^{\infty}&\Bigg[\exp\left\{\frac{im(x_j-x_r-2kN\mu_0)^{2}} {2\hbar(t-t_0)}\right\} \notag\\
&-\exp\left\{\frac{im(x_j+x_r+2kN\mu_0)^{2}} {2\hbar(t-t_0)}\right\}\Bigg]\,. \label{BoxSchProp1}
\end{align}
Or from (\ref{BoxPropagator}), expanding
\[
z\left[1-\cos\left(\frac{n\pi \mu_0}{L}\right)\right]=\frac{\hbar(t-t_0)}{m\mu^{2}_0}\left[\frac{1}{2}\left(\frac{n\pi \mu_0}{L}\right)^2+\dots\right],
\]
\begin{equation} \label{BoxSchProp2}
\lim_{\substack{N\rightarrow\infty,\,\mu_0\rightarrow 0 \\ N\mu_0=L}}\frac{k_\text{Box}(x_j,t;x_r,t_0)}{\mu_0}=\frac{2}{L}\sum^{\infty}_{n=1}\sin\left(\frac{n\pi x_j}{L}\right)\sin\left(\frac{n\pi x_r}{L}\right)\exp\left\{-i\frac{n^2\pi^2\hbar (t-t_0)}{2mL^2}\right\}.
\end{equation}
Notice, however that the series on the right hand side of  (\ref{BoxSchProp1}) or (\ref{BoxSchProp2}) does not converge. As is well known \cite{Fulling}, the Schr\"odinger propagator $k^\text{Sch}_\text{Box}$ for a particle in a box is only defined distributionally, that is, only through the integral formula
\begin{align} 
\Psi(x,t)&=\int_{-L/2}^{L/2}k^\text{Sch}_\text{Box}(x,t;y,t_0)\Psi(y,t_0)dy \notag\\
&:=\frac{2}{L}\sum^{\infty}_{n=1}\int_{-L/2}^{L/2}\sin\left(\frac{n\pi x}{L}\right)\sin\left(\frac{n\pi y}{L}\right)\exp\left\{-i\frac{n^2\pi^2\hbar (t-t_0)}{2mL^2}\right\}\Psi(y,t_0)dy. \notag
\end{align}
which gives a convergent series for \textquoteleft sufficiently well-behaved' initial wave packet $\Psi(y,t_0)$.\footnote{One may also obtain a convergent series by \textquoteleft analytic continuation' to the lower half of the complex plane, that is by giving $t$ a small negative imaginary part $t\to t-i\epsilon$. The convergent series defines a Jacobi theta function, whose properties may be used to study some singular features of the propagator \cite{Fulling}.}

In contrast, as a consequence of discreteness, the polymer propagator expressions (\ref{BoxPropagator}) and (\ref{BoxPropagatorIm}) give well defined functions for arbitrary $N$ and $\mu_0>0$. The limits (\ref{BoxSchProp1}) and (\ref{BoxSchProp2}) however, have to be taken in the distributional sense.

\section{Discussion} \label{DISCUSSION}
We have constructed the propagators in the polymer representation of quantum mechanics for the hamiltonians of the free particle on the real line and a particle in a box. The propagators are given by expression (\ref{FreePropagator}) for the free particle, and expression (\ref{BoxPropagator}) or equivalently (\ref{BoxPropagatorIm}) for the particle in a box. In our analysis we verified directly that the propagators give consistent quantum evolution determined by conditions (\ref{PolymerEvolution})-(\ref{GreensFn}), showing explicitly how standard results and constructions in the Schr\"odinger representation arise in the polymer framework. In section \ref{CONTINUUMLIMIT}, we have carefully defined the limit $\mu_0\to 0$ of these expressions, which then reduce to the standard Schr\"odinger propagators (\ref{FreeSchrodingerPropagator}) and (\ref{BoxSchProp2}). This is an important check if one is to consider the polymer description as the more fundamental one, taking into account  an underlying discreteness of space and time. In this case the discreteness is associated with the free parameter  of the polymer theory $\mu_0$, determining the lattice spacing on the real line on which evolution takes place. Another interesting line of research regarding propagators in Polymer Quantum Mechanics and their continuum limit is the renormalization group approach developed in \cite{Corichi1} in which $\mu_0$ is a renormalization parameter.

As for the polymer free field propagator  \cite{Hossain,GarciaChungX} further analysis is underway along the lines of the present work in regard to the fermonic case \cite{GarciaChungY} and which will be reported elsewhere \cite{GarciaChungZ}.

The propagator formulation for quantum evolution does not necessarily play a prominent role in the analysis of non-relativistic quantum mechanical systems. Nevertheless, explicit calculations of propagators for free particles and other simple mechanical systems, provide an overall consistency check of the polymer framework. As a toy model, our results also show explicitly in this reduced context how one may arrive at standard constructions in quantum theory from the non-standard and unitarily inequivalent representations that emerge from a loop quantization. Future investigations may not only provide insights into the application of the polymer description in quantum field theory but also make connection with some of the spinfoam-like formulations in Loop Quantum Cosmology \cite{AshtekarCampiglia}.  Whereas the kinematical aspects of the polymer description are fairly well understood and under control, dynamics still require further attention  even in successful physical theories like LQC.  With our explorations on propagators, we hope to move a step forward towards a better understanding of dynamics in the polymer setting.


\section{Acknowledgments}
Partial support is acknowledged from NSF-CONACyT Grant: Strong Back Reaction Effects in Quantum Cosmology. EFG was partially supported by SNI-III research assistant 14585 7733. JDR was supported by postdoctoral scholarship UAM-Iztapalapa.

\bibliographystyle{elsarticle-num}
\bibliography{polymerPropagatorsPreprint}

\end{document}